# A study of Type B uncertainties associated with the photoelectric effect in low-energy Monte Carlo simulations

Christian Valdes-Cortez[1,2], Iymad Mansour[3], Mark J. Rivard[4], Facundo Ballester[1,5], Ernesto Mainegra-Hing[6], Rowan M. Thomson[3], Javier Vijande[1,5,7]

[1] Departamento de Física Atómica, Molecular y Nuclear, Universitat de Valencia (UV), Burjassot, Spain
[2] Nuclear Medicine Department, Hospital Regional de Antofagasta, Chile
[3] Department of Physics, Carleton Laboratory for Radiotherapy Physics, Carleton University, Ottawa, Canada
[4] Department of Radiation Oncology, Alpert Medical School of Brown University, Providence, RI, USA
[5] Unidad Mixta de Investigación en Radiofísica e Instrumentación Nuclear en Medicina (IRIMED), Instituto de Investigación Sanitaria La Fe (IIS-La Fe)—Universitat de Valencia (UV), Valencia, Spain
[6] National Research Council Canada, 1200 Montreal Road, Ottawa, Canada
[7] Instituto de Física Corpuscular, IFIC (UV-CSIC), Burjassot, Spain

**Abstract**

**Purpose:** To estimate Type B uncertainties in absorbed-dose calculations arising from the different implementations in current state-of-the-art Monte Carlo codes of low-energy photon cross–sections (< 200 keV).

**Methods:** Monte Carlo (MC) simulations are carried out using three codes widely used in the low-energy domain: PENELOPE-2018, EGSnrc, and MCNP. Three dosimetry-relevant quantities are considered: mass energy-absorption coefficients for water, air, graphite, and their respective ratios; absorbed dose; and photon-fluence spectra. The absorbed dose and the photon-fluence spectra are scored in a spherical water phantom of 15 cm radius. Benchmark simulations using similar cross–sections have been performed. The differences observed between these quantities when different cross–sections are considered are taken to be a good estimator for the corresponding Type B uncertainties.

**Results:** A conservative Type B uncertainty for the absorbed dose ($k = 2$) of 1.2% – 1.7% (<50 keV), 0.6% – 1.2% (50 – 100 keV), and 0.3% (100 – 200 keV) is estimated. The photon-fluence spectrum does not present clinically relevant differences that merit considering additional Type B uncertainties except for energies below 25 keV, where a Type B uncertainty of 0.5% is obtained. Below 30 keV, mass energy-absorption coefficients show Type B uncertainties ($k = 2$) of about 1.5% (water and air), and 2% (graphite), diminishing in all materials for larger energies and reaching values about 1% (40 – 50 keV) and 0.5% (50 – 75 keV). With respect to their ratios, the only significant Type B uncertainties are observed in the case of the water-to-graphite for energies below 30 keV, being about 0.7% ($k = 2$).

**Conclusions:** In contrast with the intermediate (about 500 keV) or high (about 1 MeV) energy domains, Type B uncertainties due to the different cross–sections implementation cannot be considered subdominant with respect to Type A uncertainties or even to other sources of Type B uncertainties (tally volume averaging, manufacturing tolerances, etc.). Therefore, the values reported here should be accommodated within the uncertainty budget in low-energy photon dosimetry studies.

I.  Introduction

Advanced algorithms for brachytherapy absorbed dose calculations in commercial treatment planning systems (TPSs) are becoming common. This situation opens an interesting and challenging scenario, where accuracy improvements may be accompanied by departures from inter-clinic uniformity. In this context, a joint Working Group on Model-Based Dose Calculation Algorithms in Brachytherapy (WGDCAB) was created by the American Association of Physicists in Medicine (AAPM), the European Society for Radiotherapy and Oncology (ESTRO), and the Australasian Brachytherapy Group (ABG) to develop new methods and tools for standardized clinical commissioning. Specifically, the WGDCAB is charged with developing test case plans and engaging vendors to promote uniformity of clinical practices, building on the recommendations of the joint AAPM/ESTRO/ABG Task Group 186 report (Beaulieu *et al* 2012).

The first test cases delivered were made for $^{192}$Ir-based treatment planning systems, a radioisotope with an average photon energy of about 350 keV (Perez-Calatayud *et al* 2012), where typical Type B uncertainties in the Monte Carlo (MC) simulations of less than 0.5% (Ballester *et al* 2015, Ma *et al* 2017) were reported. Since 2015, TPSs for brachytherapy applications in the energy range below 50 keV have been developed, e.g., electronic brachytherapy (Valdivieso-Casique *et al* 2015) and COMS eye plaques (Morrison *et al* 2018). Therefore, the WGDCAB has endeavored to generate test cases for these clinical sites using state-of-the-art MC simulations. Unfortunately, the relatively large uncertainties affecting some of the cross–sections involved below 100 keV may play a relevant, even dominant, role in the Type B uncertainty budget (Andreo *et al* 2012). At low photon energies, the photoelectric effect is the most important interaction in terms of the energy transferred to secondary electrons. For that reason, it is hardly surprising that different implementations of the photoelectric libraries in MC systems account for most of the differences observed between MC codes (Andreo *et al* 2012). In this regard, the possible use of Pratt's renormalization screening approximation (PRSA) (Pratt 1960) has been extensively discussed in the literature (Sabbatucci and Salvat 2016, Seltzer *et al* 2014).

To explore these issues, Ye *et al.* (Ye *et al* 2004) simulated the absorbed dose to water delivered by an isotropic point-like source emitting mono-energetic photon beams between 10 keV and 200 keV. The source was at the center of a cylindrical water phantom of 15 cm radius and 20 cm height. The simulations were performed with PENELOPE-2001 and MCNP4 MC codes. MCNP4 used two different photon cross–section libraries, the DLC-200 and the DLC-146. These results were compared with those obtained with EGS4 in (Luxton and Jozsef 1999). The authors concluded that PENELOPE-2001 presented a better agreement with MCNP4/DLC-146 (within Type A uncertainties of about 5%) than EGS4, while MCNP4/DLC-200 showed significant differences between 10 keV and 60 keV and depths smaller than 6 cm.

Andreo *et al.* (Andreo *et al* 2012) carried out a systematic analysis of mass energy-absorption coefficients and their ratios for air, graphite, and water for photon energies between 1 keV and 2 MeV. Possible differences between NIST-XCOM (Hubbell and Seltzer 2004) and a modified PENELOPE-2011 version incorporating PRSA in the photoelectric cross–sections were explored, reporting expanded uncertainties on the mass-absorption coefficients of about 2.5% for energies up to 30 keV, decreasing gradually to tenths of a percent for 100 keV. With respect to the water-to-air mass energy-absorption coefficient ratios, the uncertainties reported were almost negligible except in the energy region below 4 keV where they reach values up to 1%.

However, the use of the PRSA has been controversial due to inconclusive evidence regarding the agreement between calculated and measured cross–sections. The ICRU Report 90 (Seltzer *et al* 2014) made a detailed technical and historical review of the photoelectric cross–sections calculations and their effect on the mass energy-transfer and mass energy-absorption coefficients. One of the main issues addressed is the use of PRSA. Although this correction was already included in the early 80s in some physical libraries, *e.g.* (Hubbell and Gimm 1980), it was deprecated shortly after because its worsens agreement with experimental data for photon energies lower than 1 keV. PRSA libraries were also removed for energies above 1 keV despite the fact that there was no experimental evidence supporting it at that time. More recently, Buhr *et al.* (Buhr *et al* 2012) performed measurements of the mass energy-absorption coefficients in air, for energies between 3 keV and 60 keV, with uncertainties within 1% ($k = 2$). A better agreement between the experimental data and the calculated values was found when PRSA was considered. However, Kato *et al.* (Kato *et al* 2010) carried out measurements of the mass attenuation coefficient in air for photons between 2 keV and 4 keV. The data reported in this publication, with uncertainties within 1% ($k = 2$), presents better agreement without PRSA. The ICRU Report 90 refrains from issuing any official recommendations in this regard.

Mainegra-Hing *et al*. (Mainegra-Hing 2019) compared half-value layer (HVL) and air attenuation correction ($A_{att}$) of a free-air chamber measurements with their simulated values for beams between 10 and 80 kV. The simulations were performed with EGSnrc (Kawrakow *et al* 2019) using different physical libraries, including photoelectric cross–sections with and without PRSA. With PRSA, maximum differences of 0.75% in $A_{att}$ (corresponding to a 25% difference in the air attenuation cross–section) and 35% in the HVL are reported. On the other hand, when PRSA is applied, the maximum differences reported are reduced to 0.3% and 8% for $A_{att}$ and HVL, respectively.

Basaglia *et al*. (Basaglia *et al* 2020) took a different approach based on using different statistical metrics on the photoelectric cross–sections evaluated using different models. No dosimetry calculations were performed. The experimental data include elements up to $Z = 92$ and photon energies between 100 eV and 1 MeV. According to their analysis, EPDL97 (Cullen et al 1997) cross–section tabulations, based on Scofield calculations (Scofield 1973) without PRSA, gave better agreement with respect to total and K-shell photoelectric experimental cross–sections. According to the authors, modern photoelectric libraries incorporating PRSA present statistically significant differences with respect to the experimental data.

Therefore, there is a lack of consensus regarding the proper description of the photoelectric effect in MC simulations for medical physics. Hence, until this issue is resolved, any possible difference arising from them should be included in the Type B budget uncertainty. For this reason, the present work estimates these Type B uncertainties in absorbed dose, mass energy absorption coefficients, and fluence spectra. To do so, MC simulations are carried out using three state-of-the-art MC codes: PENELOPE, EGSnrc, and MCNP6 for a selected set of energies below 200 keV using a simple geometry commonly used in brachytherapy studies.

## II. Material and methods

II.A. Monte Carlo study

The simulations performed consist of the transport of ten mono-energetic photon beams (5 keV, 10 keV, 15 keV, 35 keV, 45 keV, 55 keV, 75 keV, 100 keV, 150 keV, and 200 keV) emitted isotropically from a point-like source located at the center of a sphere of 15 cm in radius. Water has been recommended by the IAEA Code of Practice TRS-398 (Andreo *et al* 2000) and AAPM TG-253 (Fulkerson *et al* 2020) as the reference medium for low energy kV photon beams. Therefore, liquid water with the composition recommended by ICRU Report 37 (Berger *et al* 1984) and the updated mean excitation energies (78 eV) and mass density (0.998 g/cm$^3$) given by ICRU Report 90 (Seltzer *et al* 2014) is considered as the transport medium. The mean free path (MFP) in water for all the energies studied is shown in Figure 1; therefore, full scatter conditions are expected for all energies at radii less than 10 cm (Perez-Calatayud *et al* 2004).

The following observables are evaluated:

i. Mass energy-absorption coefficients in three different materials of interest in clinical dosimetry: water (same composition as the water phantom defined above), graphite (Seltzer *et al* 2014), and standard dry air (Berger *et al* 1984).

ii. Absorbed depth dose using spherical shells from $r = 0$ to $r = 5$ cm with a $\Delta r = 0.1$ cm grid. Due to the energies considered, charged particle equilibrium is guaranteed. Additionally, radiation yield is negligible, and therefore collisional kerma is scored as a surrogate for absorbed dose. For PEN18 this has been done by performing event-by-event analog simulations where electron transport has not been considered and its energy assumed to be deposited on the spot, i.e., an infinite energy cutoff for electrons is considered while retaining a 1 keV energy cutoff for photons. MCNP and EGSnrc consider the same energy cutoffs but collisional kerma is scored using a track-length estimator (Williamson 1987).

iii. Photon-fluence spectrum is obtained for 15 keV, 45 keV, 100 keV, and 200 keV at ¼, ½, and ¾ of their corresponding MFP depths (between 0.2 cm to 5.5 cm). They are scored in spherical shells with a thickness of $\Delta r = 0.1$ cm.

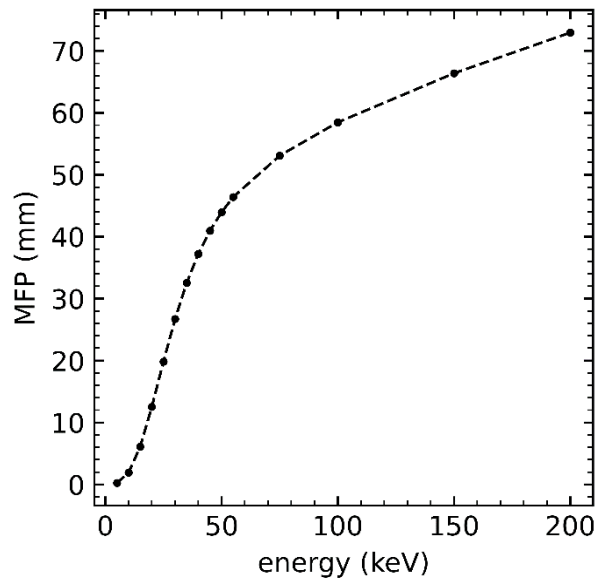

Fig 1. Mean free path (MPF) in water as a function of the photon energy considered in this benchmark.

During the last decades, the standard for MC codes was the independent electron model based on the Dirac–Hartree–Fock–Slater (DHFS) self-consistent potential. This approximation considers a single electron in a central potential, interacting with a photon without taking into account the influence of the electrons in other orbitals. This approach facilitates the calculations, but it is expected to deliver less accurate results than a more elaborate atomic model such as multi-configuration Dirac-Fock (MCDF), which involves a non-local potential, different for each sub-shell. Those considerations complicate the calculations needed, making them unavailable for a MC simulation of a realistic experimental setup (Sabbatucci and Salvat 2016). To solve this problem, Pratt's renormalization screening approximation is considered. This method corrects the DHFS cross–sections by the ratio of the electronic densities calculated with MCDF and DHFS at the limit when the radius tends to zero. All the MC systems discussed in this work use the independent electron model corrected, i.e., not using PRSA. For this reason, we adopt the following terminology: "DHFS" is used to refer to all data obtained without the PRSA; "MCDF" is used to refer to all data obtained using the PRSA. Results obtained with (MCDF) and without (DHFS) considering the PRSA are compared to illustrate the effect in the three quantities mentioned above of the photoelectric cross–section implementation. For that purpose, EGSnrc (DHFS), EGSnrc (MCDF), MCNP (DHFS), PENELOPE-2018 (DHFS), and PENELOPE-2018 (MCDF) simulations have been performed for all energies and observables.

II.B. Monte Carlo codes

Three different MC codes have been used: i) the PENELOPE-2018 MC system (henceforth denoted PEN18) (Salvat 2019) together with the penEasy v. 2019-09-21 code (Sempau *et al* 2011), ii) EGSnrc-2017 (Kawrakow *et al* 2019) with egs_brachy-2017.10.02 (Chamberland et al 2016) (EGSnrc in the following), and iii) MCNP6 (version 1.0) henceforth denoted as MCNP (Goorley *et al* 2012).

PENELOPE simulates electron and photon transport from 50 eV to 1 GeV, having been extensively used in the description of low energy phenomena. PEN18 obtains the photoelectric libraries from PHOTACS (Sabbatucci and Salvat 2016), a Fortran program that calculates subshell cross–section for arbitrary atomic potentials. PHOTACS uses the DHFS self-consistent potential supplemented by the Pratt's renormalization screening approximation. The user can choose to exclude this correction (Salvat 2019).

The Rayleigh scattering cross–sections are calculated using non-relativistic perturbation theory, obtaining the atomic form factors from EPDL97 (Cullen *et al* 1997). Compton interactions use the relativistic impulse approximation, which takes into account both binding effects and Doppler broadening (Ribberfors 1983). Furthermore, PEN18 explicitly simulates the emission of characteristic x-rays, Auger and Coster-Kronig electrons that result from vacancies produced in K, L, M, and N shells, using transition probabilities extracted from the Evaluated Atomic Data Library (EADL) (Perkins *et al* 1991). The energy of the x-rays published in the EADL was updated, when available, with the K and L shell transitions from Deslattes *et al.* (Deslattes *et al* 2003), and the M shell transitions from Bearden (Bearden 1967). Other transition energies are calculated from the energy eigenvalues of the Dirac-Hartree-Fock-Slater equations for neutral atoms (Perkins *et al* 1991).

EGSnrc is an MC toolkit for the simulation of the coupled transport of charged particles ($e^-$ and $e^+$) and photons in the energy range between 1 keV and 100 GeV in arbitrary media and geometries. Although a set of *default* transport and interaction parameters have been judiciously selected, users have great flexibility in choosing the physical models from different EGSnrc options or user-supplied data. Cross–sections from the NIST library XCOM (Berger *et al* 2010) are used by *default* for photon interactions in the energy range of interest for brachytherapy, except for incoherent scattering which, to account for binding effects and Doppler broadening, is modeled according to the relativistic impulse approximation (Ribberfors 1975, Ribberfors and Berggren 1982) using shell-wise Hartree-Fock Compton profiles (Bigg *et al* 1975). However, the use of total Compton cross–sections from the XCOM library or any user-supplied compilation can be requested via the input file. In its current *default* implementation, EGSnrc models the atomic photoelectric effect using unnormalized photoelectric cross–sections (Scofield J H 1973) by sampling interactions with the K and L shells followed by atomic relaxations using transition probabilities from the EADL library (Perkins *et al* 1991). Initial vacancies beyond the $L_{III}$ shell are not explicitly modeled, and the photo-electron receives the entire incident photon energy. This approach partially accounts for spreading out the binding energy of outer shells around the interaction point. EGSnrc has the option to use photoelectric cross–sections incorporating PRSA (Sabbatucci and Salvat 2016). If this option is selected, photoelectric interactions are sampled from any shell with binding energies above 1 keV. This effectively translates into interactions with all atomic shells out to $N_{IV}$ for Einsteinium (Z=99) for which the binding energy is around 1 keV. The user can also use the *legacy* implementation in which interactions with the K and L shells are sampled explicitly while interactions with the N and M shells are accounted for in an average manner (Kawrakow *et al* 2019). Simulation of coherent scattering is based on the independent atom approximation whereby form factors for molecules are obtained from atomic form factors (Hubbell and Gimm 1980). The user can supply their custom atomic or molecular form factors for determining Rayleigh cross–sections. Users can switch from the *default* photon cross–sections to using the photon cross–sections from the EPDL97 library (Cullen *et al* 1997) as well as from any other photon cross section library as long as it follows the proper format and naming convention.

The MCNP6 radiation transport code (version 1.0) is a general purpose MC radiation transport code that tracks nearly all particles as well as photons and electrons at energies between 1 keV to 1 GeV for electrons, and 1 eV to 100 GeV for photons. Photon cross sections, form factors, and fluorescence data are all derived from the ENDF/B-VI.8 data library for photon energies up to 1 eV (CSEWG-Collaboration 2015). In addition to extending some pre-existing data to lower energies, the ENDF/B VI.8 databases also includes subshell photoelectric cross sections. The photoelectron directions are obtained by means of an algorithm (Seltzer 1988) relying on precomputed tables based on work by Fischer and by Sauter (Fischer 1931, Sauter 1931).

Table 1 summarizes the details of the MC simulations performed in this work following the recommendation of AAPM TG-268 (Sechopoulos *et al* 2018).

*Type B uncertainties associated with photoelectric effect simulations*

Table 1: Summary of the main characteristics of the Monte Carlo simulations used in this work.

| Item | PENELOPE | EGSnrc | MCNP |
|---|---|---|---|
| Code | PENELOPE-2018 (Salvat 2019), penEasy (v. 2019-09-21) (Sempau et al 2011) | EGSnrc-2017 (Kawrakow et al 2019), egs_brachy- 2017.10.02 (Chamberland et al 2016) | MCNP6 (version 1.0) (Goorley et al 2012) |
| Validation | Previously validated (Valdes-Cortez et al 2019, Croce et al 2012, Ballester et al 2015) | Calculations of TG-43 parameters, brachytherapy treatment simulations (Chamberland et al 2016, Safigholi et al 2020) | Previously validated (Rivard et al 2010, 2006) |
| Timing | Dose to water, photon-fluence spectra, and $(\mu_{en}/\rho)_w$: On average $2 \times 10^{11}$ histories 412 h (CPU time) per simulation. | Dose to water, photon-fluence spectra: $5 \times 10^{10}$ to $1 \times 10^{13}$ histories and 600 h (CPU time) per simulation. $(\mu_{en}/\rho)_w$: $1.2 \times 10^{11}$ histories and 118 h (CPU time). | Dose to water, photon-fluence spectra: On average $5 \times 10^{10}$ histories and 2,958 h (CPU time) per simulation. |
| Source description | Photon source model: isotropic point-like source. Monoenergetic beams of 5, 10, 15, 20, 25, 30, 35, 40, 45, 50, 55, 75, 100, 150, and 200 keV. | | |
| Cross–sections | Photoelectric calculated with PHOTACS (Pratt's renormalization screening) (Sabbatucci and Salvat 2016). Rayleigh scattering using non-relativistic perturbation theory. Compton uses a relativistic impulse approximation (Ribberfors 1983). Atomic relaxation using the EADL transition probabilities (Perkins et al 1991, Deslattes et al 2003). | XCOM photon cross–section database and MCDF-XCOM photon cross–section database (Sabbatucci and Salvat 2016); Rayleigh scattering and electron impact ionization (Kawrakow et al 2019); Atomic relaxation using the EADL transition probabilities (Perkins et al 1991, Deslattes et al 2003, Bearden 1967) | Photon cross sections, form factors, and fluorescence data from the ENDF/B-VI.8 data library (CSEWG-Collaboration 2015). The photoelectrons characterized using an algorithm relying on precomputed tables (Fischer 1931, Sauter 1931, Seltzer 1988). |
| Transport parameters | Photon cutoff = 1 keV<br>Electron cutoff = $\infty$ (not transported) | | |
| Variance reduction | None | Track-length estimation is used for dose (kerma approximation), fluence and energy spectrum calculations (Chamberland et al 2016, Goorley et al 2012) | |
| Scored quantities | Absorbed dose calculated using collision kerma approximation, fluence, mass energy-absorption coefficients | | |
| Statistical uncertainties | History-by-history calculation of statistical uncertainties. $\leq 0.1\%$ ($k = 2$) | | |
| Post-processing | None | | |

II.C. Uncertainty evaluation

*Type B uncertainties associated with photoelectric effect simulations*

The standard protocol for evaluating uncertainties is the Guide to the expression of Uncertainties in Measurements (GUM) (Joint committee for guides In measurements 2008). This protocol supersedes the outdated terms of random and systematic errors (ill-defined and prone to confusions) in favor of Type A and B uncertainties. Type A uncertainties are those evaluated by the users by performing statistical analysis of their measurements. Type B uncertainties are those evaluated by any other means. Whichever the type of uncertainty, its evaluation is based on statistical distributions and, therefore, they should be quantified, if possible, in terms of their respective variances. For the former, MC simulations leading to mean values for dose to water and photon-fluence spectra allow us to obtain Type A uncertainties below 0.1% ($k = 2$) using standard history-by-history techniques (Walters *et al* 2002). For the second type of uncertainty, the GUM does not specify a unique protocol to determine it, offering to the user different possibilities for a given dataset based on its range of variation.

Different assumptions can be made depending on the amount of information available. If there is no specific knowledge about the possible values within an interval, one can only assume that it is equally probable for the *true* value to lie anywhere within it. That corresponds to a uniform or rectangular distribution described by the highest ($a_+$) and lowest ($a_-$) values obtained. In this case, the GUM gives the standard uncertainty ($k = 1$) as

$$u_B = \left| \frac{a_+ - a_-}{2} \right| \frac{1}{\sqrt{3}}. \tag{1}$$

If more information regarding the true value location is known, other less restrictive alternatives, Gaussian, triangular, trapezoidal, etc., may be used. In this particular case, and due to the absence of further information on the distribution, we have preferred to use the more conservative option, *i.e.*, larger uncertainties, namely to assume a uniform distribution limited by the results obtained using different MC physical libraries and Eq. (1) as a good estimator for Type B uncertainties. (Andreo *et al* 2012)

### III.  Results

In the following, we present results for mass energy-absorption coefficients, absorbed dose, and photon fluence with (MCDF data) and without (DHFS data) PRSA. Type A uncertainties are less than 0.06% for all cases (the largest value for the range considered).

III.A. Mass energy-absorption coefficients.

DHFS and MCDF mass energy-absorption coefficients have been evaluated using PEN18 for three materials of interest in clinical dosimetry (water, air, and graphite).. The DHFS to MCDF ratios for PEN18 are shown in Figure 2 (left). Differences between DHFS and MCDF mass energy-absorption coefficients exceeding Type A uncertainties can be observed in all three materials for energies lower than about 100 keV. The maximum differences observed are about 2.7% for air or water and about 4% for graphite. A detailed review of these differences can be found in Table SM4 as Supplementary Material. For energies below 30 keV, Type B uncertainties about 1.5% ($k = 2$) for water and air and 2 % for graphite have been found. For energies between 40 and 50 keV, these uncertainties are reduced up to 1% ($k = 2$), diminishing even further, less than 0.5% ($k = 2$), for energies larger than 75 keV. In Figure 2 (right), we show DHFS vs. MCDF differences in the mass energy-absorption coefficients water-to-air and water-to-graphite ratios. Differences beyond Type A uncertainties are only observed for the mass energy-absorption coefficients water-to-graphite ratio. In the case of water-to-graphite ratio (see additional information in Table SM5 in Supplementary Material), the only significant Type B uncertainties are observed for energies lower than 30 keV, being about 0.6% – 0.9% ($k = 2$).

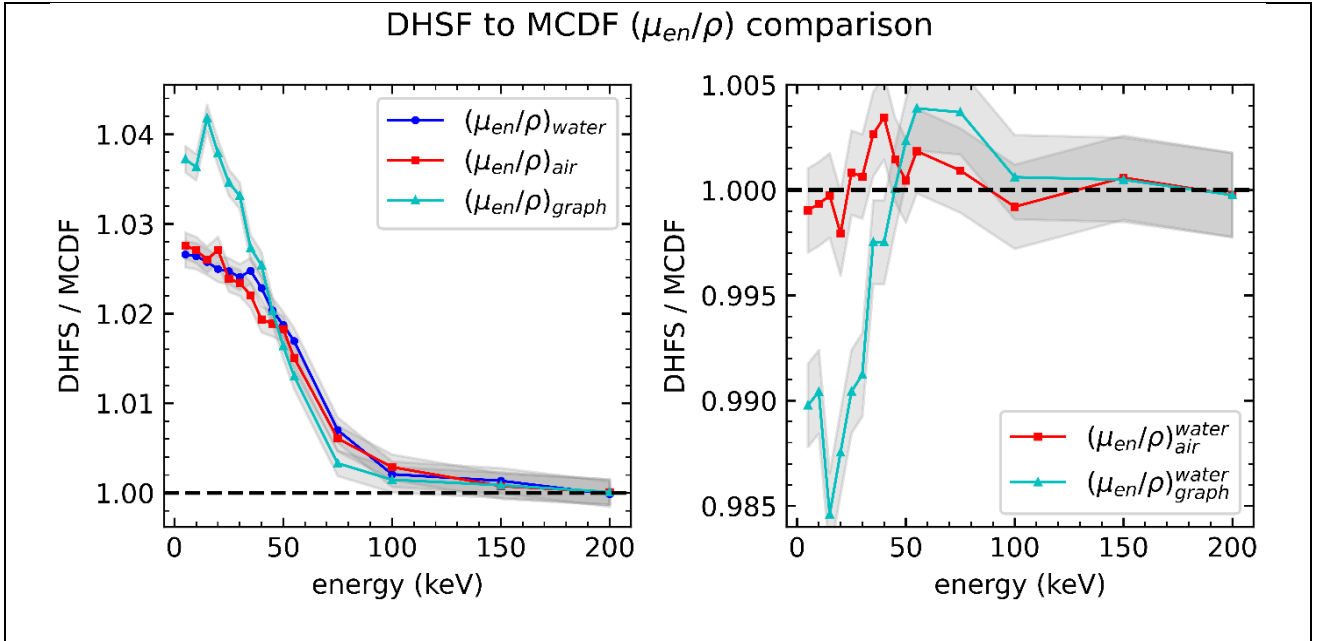

Fig 2: Mass energy-absorption coefficients comparison between DHFS and MCDF photoelectric libraries, calculated with PENELOPE-2018. Comparison of the mass energy-absorption coefficients for water, air, and graphite (left). Comparison of the water-to-material mass energy- absorption coefficient ratios (right). Grey zones represent Type A uncertainties ($k = 2$) obtained for each curve.

III.B. Absorbed dose comparisons using similar photoelectric cross–sections implementations.

Figure 3 shows the absorbed dose and their differences calculated with EGSnrc and PEN18, using a similar photoelectric implementation (DHFS). For depths smaller than 3 MFP, the agreement between the codes is within 0.5% for all energies. In the very low dose region (10 MFP), maximum differences smaller than 3% and 1.5% are observed for energies below 15 keV for MCNP (see tables SM1 in Supplementary Material) and EGSnrc, respectively, as compared with PEN18. At 5 MFP, the differences are lower than 1% and for the same range of energies. Similar results (included in the Supplementary Material, Table SM2) have been observed when comparing EGSnrc (MCDF) and PEN18 (MCDF) implementations. A complete discussion on maximum absorbed-dose differences with respect to PEN18 for each energy and depth is given as Supplementary Material.

*Type B uncertainties associated with photoelectric effect simulations*

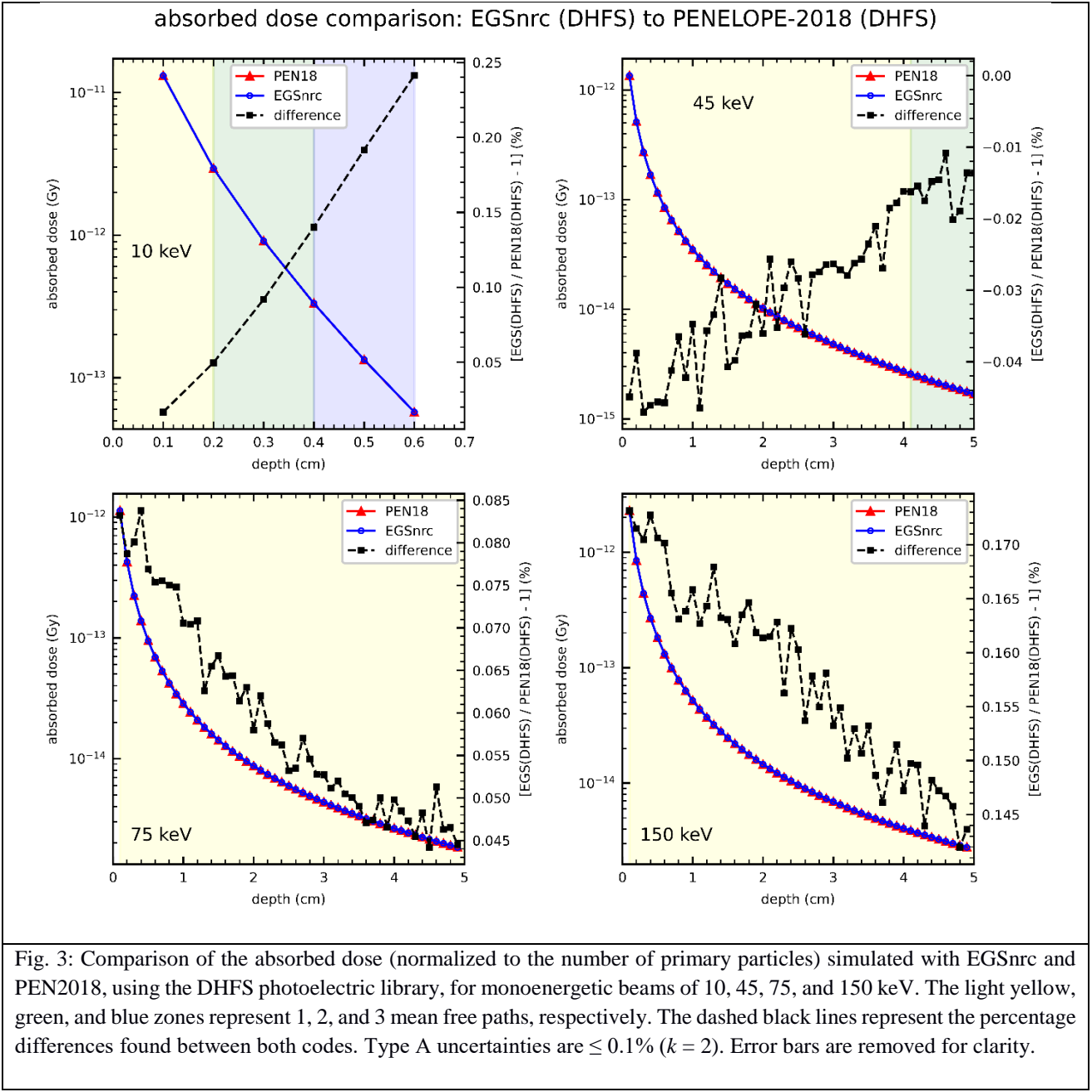

Fig. 3: Comparison of the absorbed dose (normalized to the number of primary particles) simulated with EGSnrc and PEN2018, using the DHFS photoelectric library, for monoenergetic beams of 10, 45, 75, and 150 keV. The light yellow, green, and blue zones represent 1, 2, and 3 mean free paths, respectively. The dashed black lines represent the percentage differences found between both codes. Type A uncertainties are $\leq 0.1\%$ ($k = 2$). Error bars are removed for clarity.

III.C. Absorbed dose comparisons using different photoelectric cross–sections implementations.

Figure 4 shows the absorbed dose simulated with different photoelectric cross–section implementations, DHFS vs. MCDF, together with the differences observed. For energies below 15 keV, differences about 20%, 10%, and 5% can be observed at 10, 5, and 3 MFP, respectively. At depths smaller than 2 MFP, energies between 15 and 45 keV present maximum differences between 2% and 3%. Energies between 50 and 75 keV present maximum differences between 1% and 2% for all depths. Considering depths smaller than 2 MFP and the range of energies between 15 and 55 keV, the maximum difference is at 1 mm depth (see Figure 4 for 45 keV). Finally, energies larger than 100 keV present differences lower than 0.5%. A complete report, including all energies and depths considered in this work, is included as Supplementary Material.

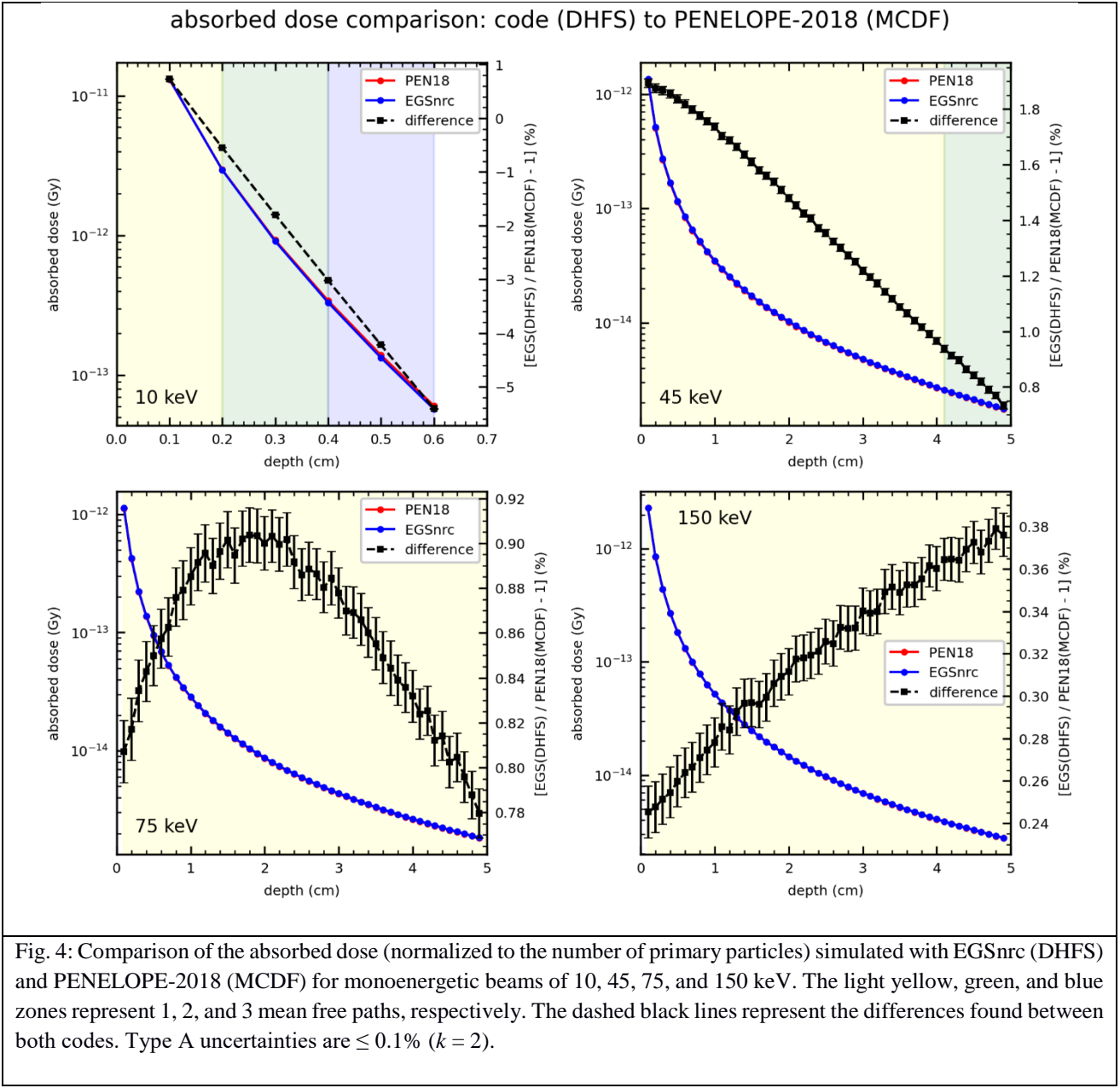

Fig. 4: Comparison of the absorbed dose (normalized to the number of primary particles) simulated with EGSnrc (DHFS) and PENELOPE-2018 (MCDF) for monoenergetic beams of 10, 45, 75, and 150 keV. The light yellow, green, and blue zones represent 1, 2, and 3 mean free paths, respectively. The dashed black lines represent the differences found between both codes. Type A uncertainties are $\leq 0.1\%$ ($k = 2$).

III.D. Photon-fluence spectra comparisons using similar photoelectric cross–sections implementations.

Photon-fluence spectra for selected energies and MFP values are shown in Figure 5 for MC codes implementing PRSA in their photoelectric cross–sections (MCDF). Differences have been evaluated as a weighted average, obtaining 0.2% (15 keV), 0.5% (45 keV), 0.3% (100 keV), and 0.1% (200 keV), hence mostly compatible with Type A uncertainties. Larger differences appear in the lowest energy regions and a set of discrete energies about 110 keV, 75 keV, 50 keV, 35 keV, and 12 keV. The latter might present maximum differences up to 15%. In both cases the photon-fluence spectra values are about four orders of magnitudes smaller than their maximum values; hence, these discrepancies are not expected to play any significant role in the determination of the absorbed dose.

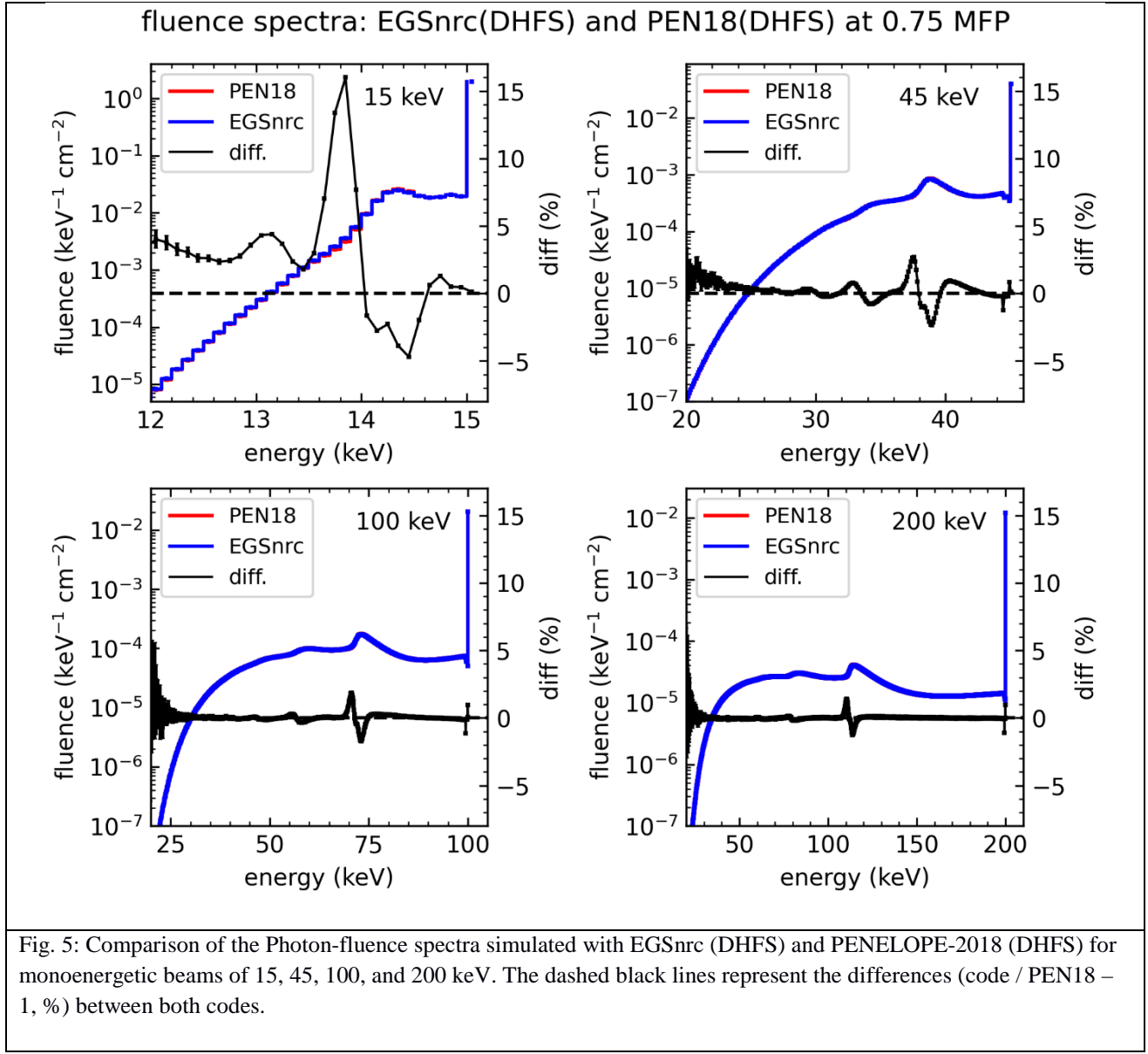

Fig. 5: Comparison of the Photon-fluence spectra simulated with EGSnrc (DHFS) and PENELOPE-2018 (DHFS) for monoenergetic beams of 15, 45, 100, and 200 keV. The dashed black lines represent the differences (code / PEN18 – 1, %) between both codes.

III.E. Photon-fluence spectra comparisons using different photoelectric cross–section implementations.

We compare in Figure 6 photon-fluence spectra for some representative MFP values for MC codes implementing different photoelectric cross–sections, EGSnrc (DHFS) and PEN (MCDF). Differences have also been evaluated as a weighted average, being 0.9%, 0.7%, 0.3%, and 0.2% for initial photon energies of 15 keV, 45 keV, 100 keV, and 200 keV respectively. Differences larger than 5% arise for photon-fluence spectra values two orders of magnitudes smaller than their maximum values, hence higher energies than in Section III. D. The same discrepancies at about 110 keV, 75 keV, 50 keV, 35 keV, and 12 keV are also present.

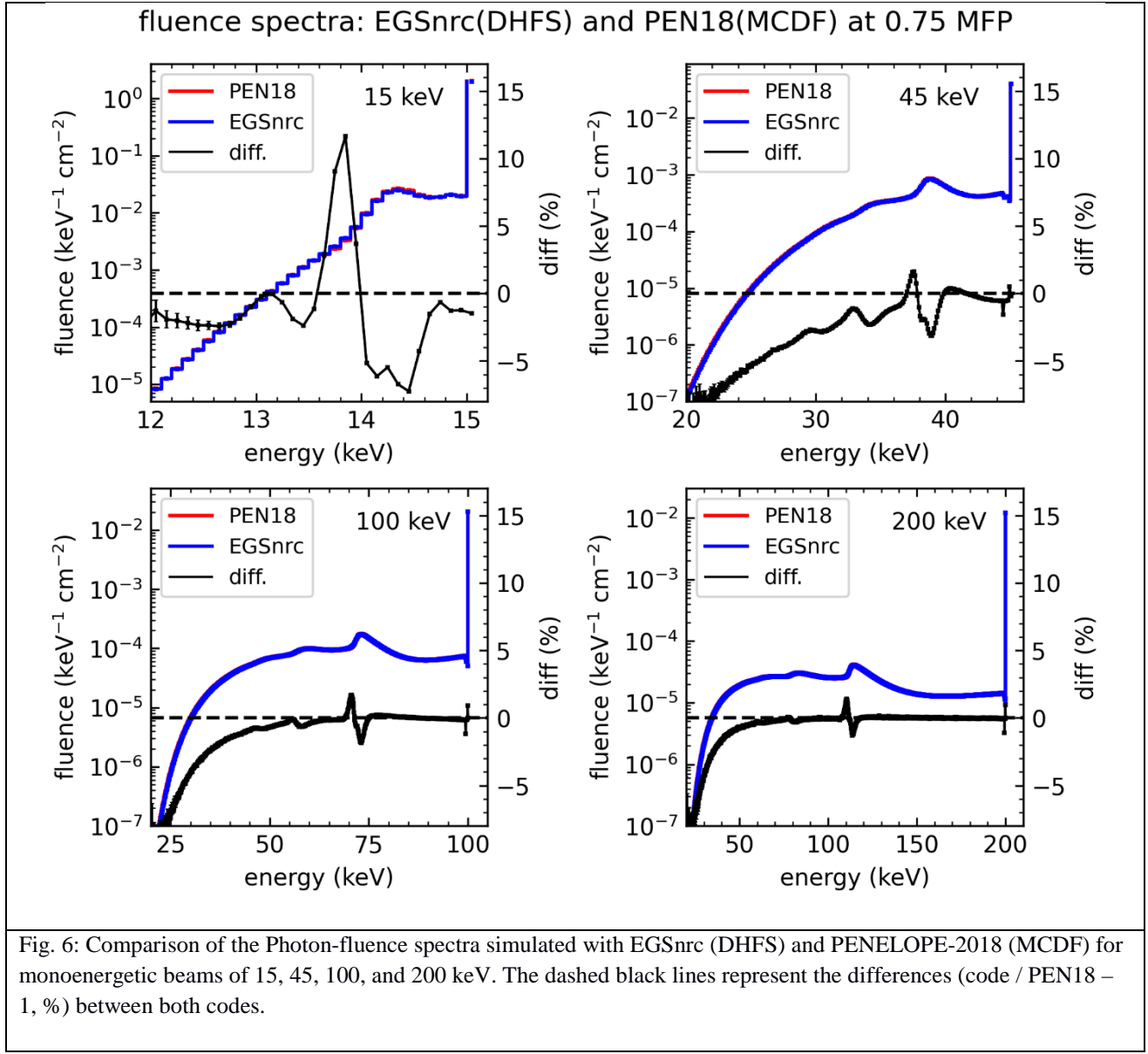

Fig. 6: Comparison of the Photon-fluence spectra simulated with EGSnrc (DHFS) and PENELOPE-2018 (MCDF) for monoenergetic beams of 15, 45, 100, and 200 keV. The dashed black lines represent the differences (code / PEN18 – 1, %) between both codes.

**IV. Discussion**

As shown above, the maximum differences observed in the mass energy-absorption coefficients are about 2.7% for air or water and 4% for graphite. It would be expected that such differences translate into similar discrepancies when evaluating the absorbed dose. However, in low-energy x-ray beams dosimetry based on ionization chamber calibrations in terms of air kerma, the absorbed dose to water is determined from measurements made free-in-air (Ma *et al* 2001, Andreo 2019). To do such air-to-water transformation, the most relevant photon-related quantities are not the mass energy-transfer coefficients for air or graphite and water, but their ratio. In turn, in this energy range, these can be replaced without any approximation by the corresponding mass energy-absorption coefficients (Andreo 2019). The general trend observed is that differences are considerably reduced when ratios are considered. It can be observed in Figure 2 (right) how, in the case of water-to-air mass energy-transfer coefficients ratio, no significant differences larger than Type A uncertainties are found. This is not the case for the water-to-graphite ratio, where differences of the order of 0.5% can still be observed for energies lower than 35 keV, reaching maximum differences of the order of 1 – 1.5% for energies ≤ 20 keV. Therefore, only for ionization chamber dosimetry involving water-to-graphite corrections below, or about, 35 keV a conservative 0.6 – 0.9% value should be considered as Type B (*k = 2*) uncertainty for cross–sections.

When evaluating the photon-fluence spectra, type B uncertainties for initial photon energies lower than 25 keV can be estimated at 0.5% due to the maximum differences observed. The only discrepancies above 25 keV are observed at a set of discrete photon energies regardless of the cross–section used. We show in Figure 7, an inset showing an example of these differences. They appear in all cases at the energies produced by Compton backscatter; namely, given an initial photon energy of 200 keV, a backscattered photon with an energy of 112.2 keV produced. This one creates a backscattered photon with an energy of 77.9 keV, and iterating one end up with similar photons at 59.7 keV, 48.4 keV, 40.7 keV, etc. As expected, the relevance of such contributions diminishes both with depth (smaller differences at ½ MFP than at ¾ MFP) due to the reduced amount of backscatter material and with the number of interactions (the effect at 77.9 keV is more reduced than at 112.2 keV). As can be seen in Figure 7, they are caused by a small redistribution of the fluence in a region of about 2-3 keV around the backscattered photon energies. These differences can be traced back to different approximations in the numerical implementation of the Compton scattering (Salvat 2019, Kawrakow *et al* 2019).

As discussed in Section II.A, collisional kerma is scored as a surrogate for absorbed dose. In doing so, the role played by the fluence spectra and the mass energy-absorption coefficients in the absorbed dose can be evaluated separately. With that purpose, we have evaluated the absorbed dose at 4 mm depth for a 15 keV beam (worst-case scenario), using the fluence-spectra simulated with EGSnrc (DHFS) and PEN18 (DHFS) and the same mass energy-absorption coefficient library from PEN18 (DHFS). The calculated absorbed doses differ by less than 0.04%. Therefore, the observed differences in the fluence spectra do not play any relevant clinical role. By repeating the same procedure but using different mass energy-absorption coefficient libraries, PEN18 (DHFS) and PEN18 (MCDF), the absorbed dose difference reaches a value of 2.6%, consistent with those reported in Section III.C.

Clinically relevant regions for photon-emitting sources below 200 keV can be restricted to distances closer than 1 – 3 cm. Hence, for the energy range studied here, these typically correspond to depths within 1 – 2 MFPs. For radioisotope-based brachytherapy, Task Group 43 Report (Rivard et al 2004) established a reference distance, 1 cm from the center of the active source along its transverse plane, located within this region. In the case of electronic brachytherapy, the typical clinical prescription depth is about 0.3 – 0.5 cm (Fulkerson *et al* 2020, Ouhib *et al* 2015). Therefore, such distances are a fundamental location where any significant uncertainty is bound to have a global effect on a clinical plan. Absorbed dose benchmarking calculations using similar cross–sections show that differences below 3 MFP depths are mostly compatible with Type A uncertainties. When different implementations are considered (see tables SM3 and SM4 in Supplementary Material), energies between 15 and 45 keV present maximum differences below 2 MFP of about 2 – 3% in the absorbed dose, diminishing up to 1 – 2% for 50 – 75 keV and 0.5% or less for 100 keV upwards. These differences are consistent with those observed for the mass energy-absorption coefficients discussed above, considering that no additional significant differences in the photon-fluence spectra above 25 keV, other than those already observed at the Compton backscattered energies, have been observed. Therefore, using the above-mentioned differences and Eq. (1), a conservative global maximum Type B (*k = 2*) uncertainty corresponding to the cross–section used of 1.2% – 1.7% (<50 keV), 0.6% – 1.2% (50 – 100 keV), and 0.3% (100 – 200 keV) should be considered in dosimetric calculations.

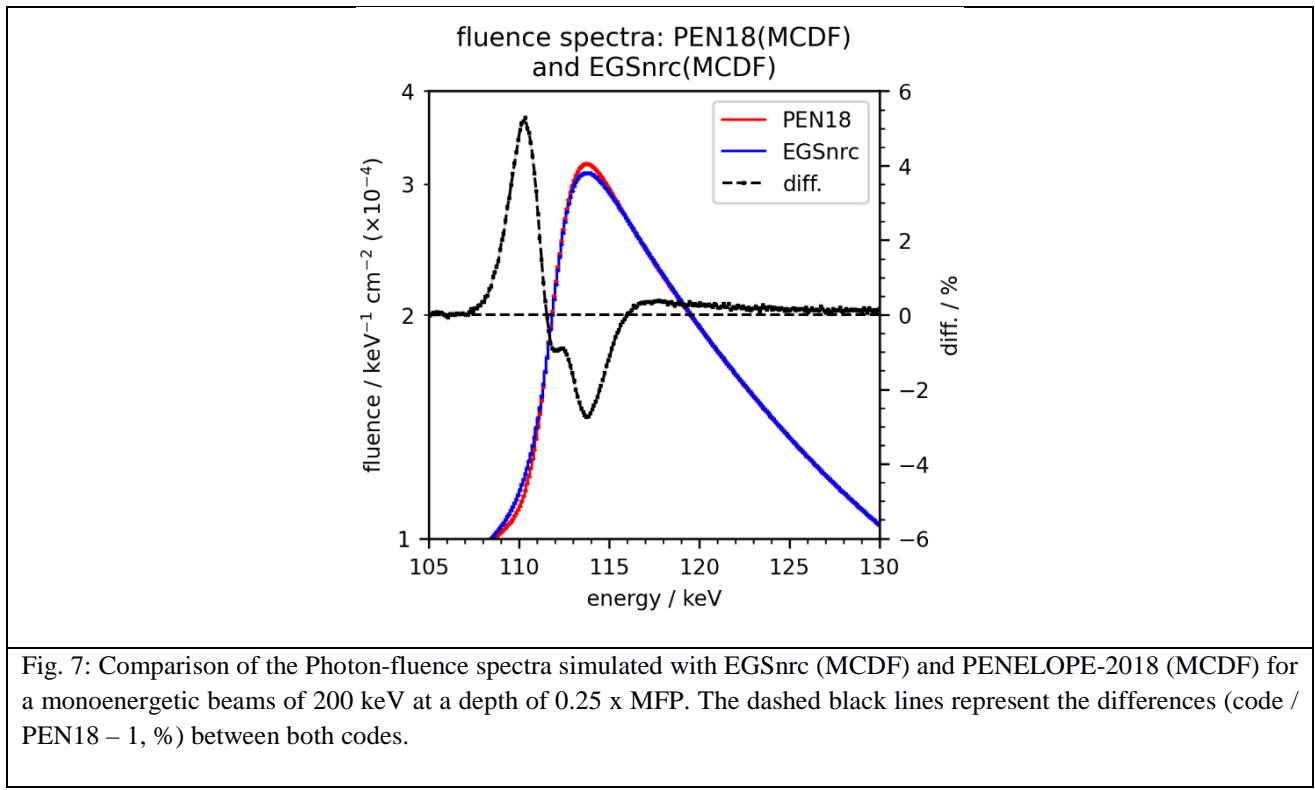

Fig. 7: Comparison of the Photon-fluence spectra simulated with EGSnrc (MCDF) and PENELOPE-2018 (MCDF) for a monoenergetic beams of 200 keV at a depth of 0.25 x MFP. The dashed black lines represent the differences (code / PEN18 – 1, %) between both codes.

To help the clinical user in establishing a more detailed Type B uncertainty for the particular low energy photon spectra used, the Type B uncertainties reported in this work are fitted using an empirical functional form (see Supplementary Material for more details):

$$u_B(E,r) = \left| a \cdot E^b r^c + u_0 \right| \tag{2}$$

where energies are given in keV and depths in cm. The fit parameters *a, b, c,* and $u_0$ are given in Table 4. This fit is valid for depths within 10 MFPs. As an illustrative example, we show in Figure 8 Type B (*k = 2*) uncertainties as a function of depth for a typical $^{125}$I spectrum, a low-energy photon-emitting radioisotope widely used in clinical brachytherapy practice.

| Table 2: Parameters *a*, *b*, *c*, and $u_0$ obtained from the fit using Eq. (4); energies in keV and distances in cm) |||
|---|---|---|
| Parameter | value | Range of energies (keV) |
| a | $-4.9227 \times 10^4$ | $5 \leq E \leq 40$ |
| b | $-2.8559$ | |
| c | $0.90255$ | |
| $u_0$ | $1.1725$ | |
| a | $-2.5808 \times 10^4$ | $40 < E \leq 200$ |
| b | $1.8780 \times 10^{-4}$ | |
| c | $2.7182 \times 10^{-5}$ | |
| $u_0$ | $2.5845 \times 10^4$ | |

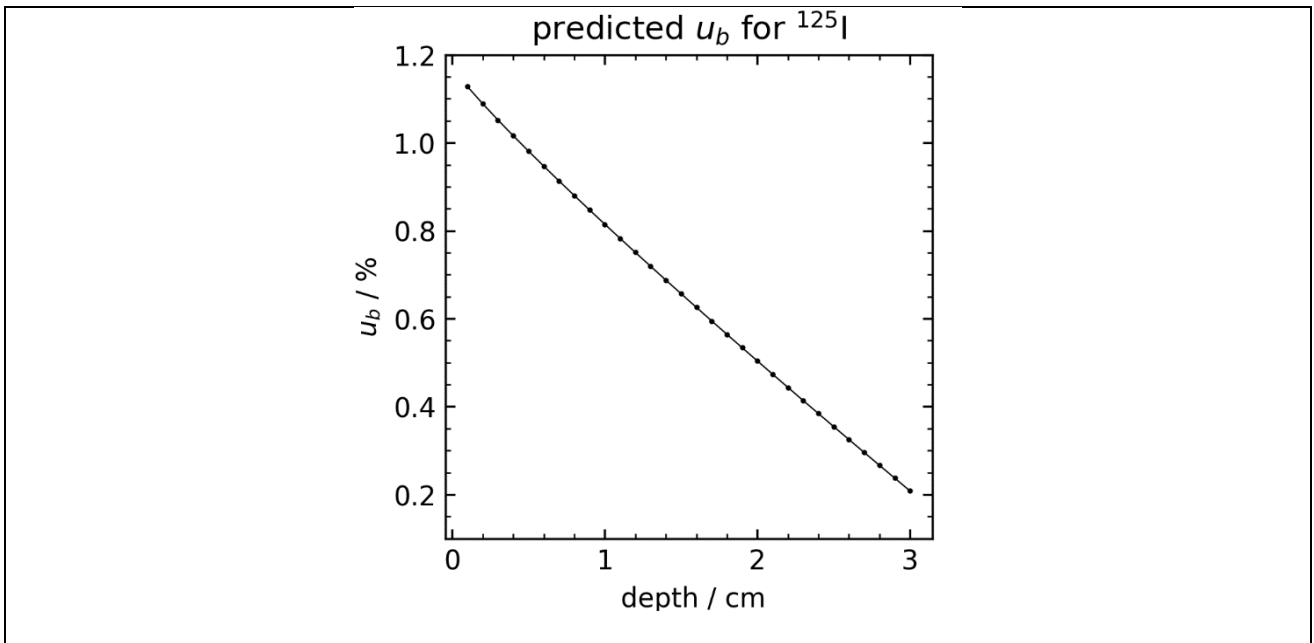

Fig. 8: Predicted Type B (*k = 2*) uncertainties (%) for $^{125}$I as a function of depth obtained with Equation 1.1 and parameters showed in Table 4.

## V. Conclusions

Type B (*k = 2*) uncertainties of about 1.5% (air and water) and 2% (graphite) have been observed for the mass energy-absorption coefficients of dosimetrically relevant materials. However, such uncertainty becomes severely reduced in the case of the mass energy-absorption water-to-air ratio. Only in the case of ionization chamber dosimetry involving water-to-graphite corrections below 35 keV a conservative Type B (*k = 2*) 0.6% – 0.9% uncertainty should be considered. The differences observed in the photon-fluence spectra are either within Type A uncertainties or not clinically relevant; hence no additional Type B uncertainty is recommended in this case with the exception of energies below 25 keV, where a Type B uncertainty of 0.5% has been estimated. With respect to the absorbed dose, a Type B (*k* = 2) uncertainty of 1.2% – 1.7% (<50 keV), 0.6% – 1.2% (50 – 100 keV), and 0.3% (100 – 200 keV) should be considered in dosimetric calculations.

Opposite to what happens in the intermediate (about 500 keV) or high (about 1 MeV) energy domains, Type B uncertainties due to the different cross–sections implementation cannot be considered subdominant with respect to Type A uncertainties or even to other sources of Type B uncertainties (tally volume averaging, manufacturing tolerances, etc.). Therefore, the values reported here should be accommodated within the uncertainty budget in low-energy dosimetry studies.

### Acknowledgments

FB and JV acknowledge funding from the Spanish Government FEDER/MCIyU-AEI and Generalitat Valenciana under grants PGC2018-101302-B and AICO/2019/132, respectively. RMT and IM acknowledge support from the Natural Sciences and Engineering Council of Canada, Canada Research Chairs program, and Compute Canada.